\documentclass[a4paper,11pt]{article}
\pdfoutput=1
\usepackage{jcappub}
\usepackage{amsmath}
\usepackage{amssymb}
\usepackage{mathtools}
\usepackage{CJK}
\usepackage{graphicx}
\usepackage{dcolumn}
\usepackage{bm}
\usepackage[dvipsnames,table]{xcolor}
\hypersetup{urlcolor=BlueViolet,
	    citecolor=Plum,
	    linkcolor=PineGreen}

\newcommand{\msun}{M_\odot}

\begin{document}
\subheader{\hfill CPPC-2021-02}

\title{Navigating the asteroid field: New evaporation constraints for primordial black holes as dark matter}

\author{Zachary S. C. Picker}
\affiliation{Sydney Consortium for Particle Physics and Cosmology}
\affiliation{Centre of Excellence for Dark Matter Particle Physics \\
 School of Physics, The University of Sydney, NSW 2006, Australia }
\emailAdd{zachary.picker@sydney.edu.au}

\abstract{Primordial black holes (PBHs) in the asteroid-mass range $(10^{-17}$--$10^{-12})~M_{\odot}$ almost entirely escape constraints as dark matter candidates. In the early universe, however, the Schwarzschild metric is no longer applicable and we must carefully consider PBH metrics with a proper cosmological embedding. In particular, the Thakurta solution stands out as perhaps the most adequate such solution. Notably, the Thakurta metric contains a time-dependent apparent horizon, proportional to the cosmological scale-factor. We show that this implies the PBHs are hotter than in the standard Schwarzschild case, and so evaporate significantly faster via Hawking radiation. By matter-radiation equality, PBHs up to masses of $\sim 10^{-12}~M_{\odot}$ have totally evaporated, and so cannot be a viable dark matter candidate.}
\maketitle

\section{Introduction}

Primordial Black Holes (PBHs)~\cite{pbh,Hawking:1971ei,Carr:1974nx,Chapline:1975ojl} are one of the oldest and perhaps most intriguing dark matter candidates, forming in the early universe from large density perturbations. Notably, the asteroid-mass range~$(10^{-17}$--$10^{-12})~M_{\odot}$ for PBH dark matter-- sitting between constraints from evaporation due to Hawking radiation~\cite{Hawking:1974sw,Carr:2009jm} and microlensing surveys~\cite{Niikura:2017zjd}-- remains almost entirely unconstrained~\cite{Montero_Camacho_2019,Green:2020jor,Carr:2020gox}. Previous bounds in this range, from femtolensing and the capturing of PBHs in stars and white dwarfs, have recently been revisited and dismissed~\cite{Montero_Camacho_2019}. Optical microlensing constraints are limited by finite source size and diffraction effects, while the rate of capture of PBHs in stars is insufficient to place strong constraints. It was thought that transiting PBHs may ignite white dwarfs and destroy them, but these bounds are challenged by recent calculations of these effects and by heavy astrophysical and theoretical uncertainties.

Schwarzschild black holes of masses below approximately $10^{-19}~\msun$ evaporate with a lifetime smaller than the age of the universe. This constitutes a `stability bound' which constrains these PBHs as dark matter candidates. Nominally, a PBH with initial mass slightly larger than this critical size would leave behind smaller remnants, which then could constitute the dark matter today. However, the mass of the remnant today depends extremely sensitively on the initial mass at formation. Since even relatively detailed calculations of PBH evolution will have less precision than this sensitivity, this constitutes a \textit{stability bound} on any theory in which there remain today PBHs smaller than $~10^{-19}~\msun$\footnote{There are presumably other constraints on such a theory, since these small PBHs would produce a significant amount of Hawking radiation which would be detectable.}. This point is not often stressed in this much detail, but will later be relevant for our arguments.

Many of the tightest constraints on PBH dark matter involve the study of PBHs in the early universe, when the black holes are embedded in the cosmological fluid. Close to the epoch of formation, however, the Hubble horizon will not necessarily be sufficiently far away from the PBH horizon that we can adequately describe such black holes using the Schwarzschild metric. There is a long history of works~\cite{McVittie:1933zz,Einstein:1945id} describing black hole-type solutions embedded in a Friedmann-Lema\^{i}tre-Robertson-Walker (FLRW) universe; see~\cite{Faraoni:2018xwo} for a recent review. However, there has not been much study or consensus on the application of these cosmological PBH solutions to their constraints as dark matter candidates. Recently, we studied the application of the Thakurta metric \cite{Thakurta}, one particularly justifiable cosmological PBH solution, to the strong constraints on LIGO-mass $(30$--$100~M_{\odot})$ PBH dark matter from early-universe binary abundance calculations~\cite{boehm2020eliminating}. The Thakurta metric posseses the interesting property of having a time-dependent quasi-local \textit{effective} mass, which is roughly proportional to the FLRW scale factor. This property made it more difficult for PBHs to decouple from the Hubble flow to form binaries, alleviating merger-rate constraints from gravitational waves detected by the LIGO-Virgo collaboration~\cite{Abbott:2016blz,Nakamura:1997sm,Ali-Haimoud:2017rtz}

In this work, we continue our study of the Thakurta metric, focusing in particular on Hawking radiation. The Thakurta PBH contains an apparent horizon which, like the mass, is roughly proportional to the scale factor. This means that at any given time in the early universe, these PBHs are significantly hotter than in the usual Schwarzschild case, and so also evaporate much more rapidly. This allows us to place tighter constraints-- by several orders of magnitude-- on asteroid-mass PBH dark matter, since PBHs which totally evaporate before today are bound by the same stability constraints as previously discussed. In fact, PBHs up to the size of approximately~$10^{-12}~M_{\odot}$ which form at their `standard' formation time (when their Schwarzschild radius coincides with the cosmological horizon) totally evaporate even before matter-radiation equality at redshift~$z_{\rm{eq}}\sim 5000$. 

This work is intended to be a `first-pass' at this sort of calculation; we will not perform the full curved-spacetime quantum field theory calculations for Hawking radiation, or delve into all the details of PBH evolution. However, we still can reach a relatively powerful conclusion straightforwardly and under conservative assumptions; Thakurta PBHs smaller than $\sim 10^{-12}~M_{\odot}$ have lifetimes smaller than the age of the universe. Of course, the choice of Thakurta metric is also an assumption we must make, although we offer a brief justification of this choice, and much of this work can apply more generally to other cosmological black hole metrics. Perhaps the most important point that we would like to stress is that, regardless of one's level of belief in the Thakurta metric, there are significant phenomenological impacts from a more careful treatment of cosmological PBHs, and these impacts have serious consequences on the viability of PBHs as dark matter candidates.

\begin{figure}[t]
    \centering
    \includegraphics[width=0.8\textwidth]{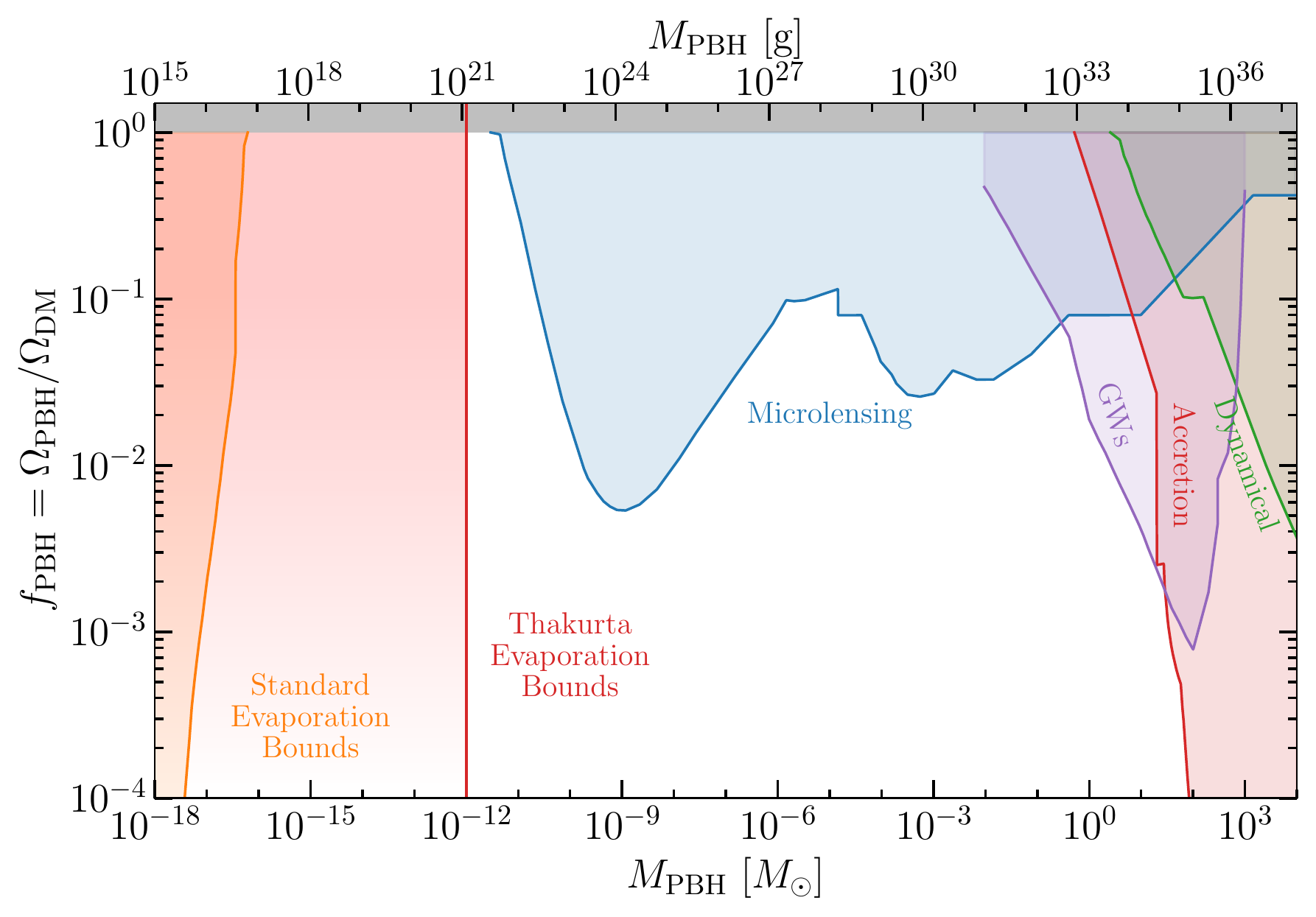}
    \caption{\label{fig:pbhconstraints}If the Thakurta metric is assumed to describe PBHs in the radiation-domination era of the early universe, then black holes which form at a mass of $\sim 10^{-12}~\msun$ evaporate with a lifetime shorter than the age of the universe. PBHs with initial mass slightly larger than this critical mass could leave behind remnants in the asteroid-mass range, but any such scenario is extremely sensitive to the value of the formation mass. As a result, we have a `stability constraint' on any such theory, and in effect, we can shift the evaporation constraints on PBH dark matter several orders of magnitude. The PBH constraint plot~\cite{bradley_j_kavanagh_2019_3538999} can be found at \url{https://github.com/bradkav/PBHbounds}.}
\end{figure}

In section~\ref{sec:metrics}, we introduce the Thakurta metric, justifying our choice from amongst its alternatives and examining some of its crucial features. In particular, we transform the metric into the Kodama time~\cite{Kodama:1979vn,Abreu:2010ru} for easier and less ambiguous study of Hawking radiation. Then in section~\ref{sec:hawk} we derive the Hawking radiation mass-loss equation for Thakurta PBHs and in section~\ref{sec:evap} we find the new evaporation bounds for PBH dark matter, before concluding in section~\ref{sec:conc}.

\section{The Thakurta metric}\label{sec:metrics}
There are a number of proposals for metrics describing central inhomogeneities embedded in FLRW backgrounds~\cite{Faraoni:2018xwo,carrera}. Many of these we addressed in greater detail in our previous work~\cite{boehm2020eliminating} so we will give just a brief summary of a few well-known solutions here. Many of these metrics are plagued with various pathologies or are only correct for specific kinds of cosmological fluid sources. 

\subsection{Survey of cosmological black holes}
One of the most well-known cosmological metrics is the Einstein-Strauss (or Swiss-cheese vacuole) solution~\cite{Einstein:1945id}, which is a stitching of a Schwarzschild solution into an FLRW metric in a ball of some radius. There are perhaps no obvious physical problems with this metric, but such a solution might be considered at least somewhat \textit{ad-hoc}, and is restricted to dust-FLRW spacetimes~\cite{carrera}. The Lema\^{i}tre-Tolman-Bondi class of metrics~\cite{Tolman:1934za,Bondi:1947fta,Joshi:2014gea} suffer from shell-crossing singularities for solutions which interpolate between the small-scale black holes and large-scale expansion, and also require dust backgrounds. Similarly, the Sultana-Deyer solution~\cite{Sultana:2005tp}, which is in some ways quite similar to the Thakurta solution, requires a dust background. The McVittie geometry~\cite{McVittie:1933zz} was an early and quite generic solution to this problem. However, the McVittie geometry does not include a radial heat flow onto the inhomogeneity, leading to a \textit{spacelike} naked singularity (and therefore divergent pressures) at the black hole horizon, unless the universe asymptotically approaches a de Sitter spacetime-- which is not applicable in the early universe, where we are interested. 

The generalized McVittie geometries~\cite{Faraoni:2007es}, however, are sourced by an imperfect fluid which is now allowed to contain a radial heat flow. For this solution, the apparent horizon becomes non-singular and and the energy density can be positive everywhere.

\subsection{The Thakurta metric in detail}
The late-time attractor solution of the entire class of generalized McVittie geometries coincides with another cosmological black hole metric, known as the non-rotating Thakurta metric~\cite{Thakurta,Faraoni:2007es}. This metric also appears to be the general-relativistic limit of a class of exact solutions of Brans-Dicke gravity with a cosmological fluid~\cite{clifton10.1111/j.1365-2966.2005.08831.x}. The Thakurta metric is valid during radiation-domination, and does not contain any of the pathologies of many of its alternatives. For these reasons, as well as its refreshing simplicity and potentially rich phenomenology, the Thakurta solution is singled out as one of the most intriguing cosmological PBH solutions.

The Thakurta solution can be written in terms of the Schwarzschild line element and cosmological scale factor $a(t)$ as

\begin{align}
    \mathrm{d}s^2 = a^2(t) \, \mathrm{d}s^2_{\rm schw} \, .
\end{align}
In terms of the `physical' (areal) radial coordinate $R=a(t)r$, cosmic time $t$ and angular coordinates $\theta, \phi$, this metric can then be written as,

\begin{equation}\label{Thakurta}
\mathrm{d}s^2=f(R)\left(1-\frac{H^2R^2}{f^2(R)}\right)\mathrm{d}t^2 + \frac{2HR}{f(R)}\mathrm{d}t \, \mathrm{d}R-\frac{\mathrm{d}R^2}{f(R)}-R^2\left(\mathrm{d}\theta^2+\sin^2\theta \,\mathrm{d}\phi^2\right)\,,
\end{equation} 
where $f(R)=1-2Gm a(t)/R=1-2Gm/r$ and $m$ is the `physical' mass of the PBH-- the mass of the black hole in the static limit, evaluated when $a=1$, which is the natural quantity with which to describe PBH dark matter at the current epoch. The Hubble expansion rate reads $H=\dot{a}/a$, with $\dot{a}$ standing for the time derivative of the scale factor. It is straightforward to show that in the small-scale limit this approaches a Schwarzschild-like spacetime and in the large-scale limit it approaches the usual FLRW metric.

The notion of the black hole mass requires a somewhat more careful definition, however, when we are in a cosmological setting. The physical mass $m$ for black holes today is commonly described using the Arnowitt-Deser-Misner (ADM) mass~\cite{Arnowitt:1959ah}, which is defined at spatial infinity. However, this notion of mass is not appropriate for the early universe, when there the cosmological horizon is not sufficiently far away from the black hole. We need a \textit{local} definition of mass; the conventional procedure is to define the quasi-local Misner-Sharp mass~\cite{Misner:1964je}, which is a measure of active mass within a given volume. This is an `effective' mass that a test object near the PBH would feel as a result of both the central object and the cosmological background. The Misner-Sharp mass can be directly calculated for the Thakurta metric:

\begin{equation}\label{MS}
m_{\mathrm{MS}}=ma(t)+\frac{H^2R^3}{2Gf(R)}~. 
\end{equation}

In the radiation-dominated era, we have that $H\sim 1/a^2$ and $R\sim a$, so that the first term quickly dominates and we can approximately treat the Misner-Sharp mass as a `comoving' mass. The fact that the Misner-Sharp mass increases with time is not a feature of any sort of accretion onto the physical mass of the PBH; it is merely an \textit{effective} mass which is a consequence of the Thakurta geometry. The Misner-Sharp mass also defines an apparent, or `trapping', horizon at $R = 2Gm_{\mathrm{MS}}$. While the definition of an event horizon requires knowledge of the entire future of the spacetime, apparent horizons can be defined locally; they are surfaces from which nothing emerges \textit{now}\footnote{For more thorough and precise surveys of black hole horizons, see Refs.~\cite{Faraoni:2018xwo,boothdoi:10.1139/p05-063}.}. The Thakurta metric is a rare case~\cite{Faraoni:2018xwo} in which the we can find an explicit solution for the apparent horizon:

\begin{align}\label{Rb}
    R_{\rm{b}} = \frac{1}{2H}\left(1-\sqrt{1-8HGma(t)}\right)~.
\end{align}

For a valid solution, we now have that $1/8> \delta \coloneqq HGma$. This is a useful small quantity we can use to simplify our calculations by keeping results to any given order in $\delta$.

\subsection{Kodama time}\label{sec:kodama}
Dynamic black holes lack an asymptotically timelike Killing vector field, which makes the computation of properties such as surface gravity (and therefore Hawking temperature) somewhat ill-defined, since there is an ambiguity in the choice of time foliation in which these processes are defined. However, there does exist a geometrically preferred vector field, given by the Kodama vector~\cite{Kodama:1979vn}. This vector defines a timelike direction, which is parallel to the Killing vector in the static limit, but does not itself naturally define a preferred time coordinate. It was later found by Abreu and Visser~\cite{Abreu:2010ru} that there was a strongly suggested preferred time coordinate, nominally called the ``Kodama time", which is the unique choice of coordinate $\tau$ that makes integral curves of $\partial_{\tau}$ coincide with integral curves of the Kodama vector. For the Thakurta metric, we can put the metric in Kodama time via the transformation,

\begin{align}
    \mathrm{d}\tau = \mathrm{d}t + \frac{H R}{f(R)}\frac{\mathrm{d}R}{1-\frac{2G m_{\rm{MS}}}{R}}~.
\end{align}
It is also useful to note that when we are in the region below the cosmological horizon (i.e., $R\ll 1/H$) but above the ``Schwarzschild" horizon ($R\gg 2Gm>2Gm_{\rm{MS}}$), the Kodama time $\tau$ approximately coincides with the cosmic time $t$. For the Schwarzschild case, the relevant observer for the black hole mass loss is at infinity; here, the observer should be within the cosmological horizon but far from the black hole, where the Kodama time and cosmic time roughly coincide. Conveniently, here we have that the scale factor $a(\tau)\sim a(t)$ and we can use the standard form of the Friedmann equations to calculate cosmological quantities. Under this transformation, the Thakurta metric becomes,

\begin{align}\label{kodamametric}
    \mathrm{d}s^2 = \left(1-\frac{2Gm_{\rm{MS}}}{R}\right)\mathrm{d}\tau^2 - \left(1-\frac{2Gm_{\rm{MS}}}{R}\right)^{-1}\mathrm{d}R^2 - R^2 \left(\mathrm{d}\theta^2+\sin^2\theta \,\mathrm{d}\phi^2\right)~.
\end{align}

This Schwarzschild-like form provides us a natural and simple way to handle dynamic black holes, making calculations of Hawking radiation significantly more straightforward and unambiguous. More generally, the norm of the Kodama time translation vector differs from the norm of the Kodama vector~\cite{Abreu:2010ru}; we find it natural here to work in coordinates where these two coincide, just as one would do in the static case where the timelike Killing vector generates translations in time.

\section{Hawking radiation}\label{sec:hawk}

Following Ref.~\cite{Abreu:2010ru}, we can now calculate the surface gravity in the Thakurta metric, to first order in $\delta$:

\begin{align}\label{kappa}
    \kappa \approx \frac{1}{2Gma} - \frac{3\delta}{Gma}~, 
\end{align}
which to zero-th order is related to the familiar Schwarzschild result by $\kappa = 2\kappa_{\rm{Schw}}/a$~\cite{Carr:2020gox}. We then make the usual assumption that the black hole temperature is given by $T = \kappa/2\pi$ and that the entropy is given by $S = A/4$, where $A$ is the surface area of the apparent horizon.

The equation describing the change in black hole mass due to Hawking radiation is most straightforwardly derived from the thermodynamic identity,

\begin{align}\label{therm}
    \frac{\mathrm{d}U}{\mathrm{d}\tau} = T\frac{\mathrm{d}S}{\mathrm{d}\tau} - P\frac{\mathrm{d}V}{\mathrm{d}\tau}~,
\end{align}
where $P$ is the pressure at the horizon, $V$ is the volume and we are using the Kodama time $\tau$ as our `most natural' time coordinate. The change in internal energy $U$ at the apparent horizon will now have contributions both from Hawking radiation, and from the fact that the apparent horizon itself is growing with time. Unlike the Schwarzschild case, the $P\mathrm{d}V$ term will not vanish; however it contributes only at higher order in $\delta$ in our final result. 

In the standard calculation, one assumes that the black hole is a static blackbody source (i.e.~one disregards the effect of the shrinking mass of the black hole) and so the left-hand side of Eq.~(\ref{therm}) is given purely by the Stefan-Boltzmann law for blackbody radiation. In our case, when $\dot{T}$ (where a dot now represents a time derivative with respect to Kodama time) is explicitly nonzero, we should first check that the black hole surface is not too far from being thermal. Explicitly we can check that the rate of temperature change is less than the rate of Hawking evaporation,

\begin{align}
\left|\dot{T}\right|<\left|\dot{U}_H\right|~,
\end{align}

where $\dot{U}_H$ is the Hawking power given by the Stefan-Boltzman law, $\dot{U}_H = -\sigma T^4 A$, and $\sigma$ is the Stefan-Boltzman constant. This inequality can be simplified to $240~Gma \lesssim 1/H$, which is always satisfied, since the black holes form when $2Gm \sim 1/H$. Since the Hawking power is always larger than the change in temperature of the black hole, we will assume that the radiation from the Thakurta black hole does indeed follow the Stefan-Boltzmann law.

We must also account for the fact that the internal energy evaluated at the horizon grows with time, since the apparent horizon itself grows with time. The internal energy was calculated in Ref.~\cite{Abreu:2010ru}; evaluated at the apparent horizon, it is simply given by $U = G R_b$. This term must be included in our calculation, and in fact is needed to cancel the time derivatives of $a$ in Eq.~(\ref{therm}). To lowest order in $\delta$, we can then replace the left-hand side of Eq.~(\ref{therm}) with:

\begin{align}\label{dudt}
    \frac{\mathrm{d}U}{\mathrm{d}\tau} = -\sigma T^4 A + 2\delta~.
\end{align}

From the Misner-Sharp mass defined in Eq.~(\ref{MS}) and apparent horizon defined in Eq.~(\ref{Rb}), it is straightforward to evaluate the temperature and entropy in terms of the physical mass $m$ and scale-factor $a$. Then, using Eq.~(\ref{dudt}), we can fully evaluate the thermodynamic identity Eq.~(\ref{therm}), solving for the rate of change of the physical mass $m$ of the black hole (noting that the $P\rm{d}V$ term can be neglected at lowest order in $\delta$). To zero-th order in $\delta$, the rate of change of mass is given by:

\begin{align}\label{dmdt}
    \frac{\mathrm{d}m}{\mathrm{d}\tau} = -\frac{1}{1920 \pi G^2 m^2 a^2}~,
\end{align}
which differs from the usual Schwarzschild rate by a factor of $8/a^2$. As we previously discussed, the difference between Kodama time $\tau$ and cosmic time $t$ will not be substantially important, since the transformation becomes trivial for smaller $m$ as long as we are within the cosmological horizon. The extra factor of $a^{-2}$, however, is enormously consequential in the early universe, allowing PBHs which form at relatively large masses to evaporate much more quickly than in the Schwarzschild case.

\section{Evaporation constraints}\label{sec:evap}
Now we have derived the fundamental equations for Hawking radiation in the Thakurta metric, we can apply these to the constraints on PBHs as dark matter candidates.

\subsection{Thakurta PBHs in the early universe}
We should first discuss when it is appropriate to treat the PBHs using the Thakurta metric. In our previous work~\cite{boehm2020eliminating}, we discussed in some depth the conditions for which a Thakurta PBH would decouple from the Hubble flow and transition into a more traditional Schwarzschild black hole. Perhaps the most natural decoupling condition occurs when the black holes are no longer dominated by the cosmological background, but instead by some local dynamics. In the early universe, this essentially requires that two PBHs encounter each other, decoupling from the Hubble flow to form a binary. We showed that this decoupling condition is actually relatively difficult to satisfy, occurring well after matter-radiation equality for the vast majority of PBHs. In addition, this condition becomes increasingly harder to satisfy as the size of the black holes decrease, so we do not have to worry about the black holes suddenly decoupling once they have evaporated a sufficient amount. 

To be conservative, and to simplify the calculations, we will assume here that all PBHs decouple at matter-radiation equality to become Schwarzschild black holes (or perhaps, something more akin to the Einstein-Strauss solution). We will also assume that accretion does not play a significant effect on the evolution of the black holes; it is known that accretion is not usually significant during radiation-domination~\cite{Carr:2020gox}, especially for masses in the asteroid- and smaller- mass range, so this assumption should not be too worrisome.

\subsection{Evaluating the critical mass}
We will then look for the critical mass PBH which is fully evaporated by matter-radiation equality at $z_{\rm{eq}}$\footnote{Usually the term `critical mass' refers to black holes which evaporate with the lifetime of the universe. Here we choose matter-radiation equality instead, in order to simplify the calculations and because the difference in mass between these two definitions will anyway be smaller than the precision of our calculations.}. Integrating Eq.~(\ref{dmdt}) between the black hole formation time and matter-radiation equality, and setting the final mass to zero, gives us a relation for the critical mass $m_*$ which fully evaporates exactly at matter-radiation equality. During radiation-domination, we can approximately use the relation for the Hubble rate in terms of redshift $H\sim H_0\sqrt{\Omega_{\rm{r}}(1+z)^4}$ to make this integration analytically tractable:

\begin{align}
    m_*^3 \sim \frac{z_{\rm{eq}}}{640 \pi G^2 H_0 \sqrt{\Omega_r}}\left(\frac{z_{\mathrm{f}}(m_*)}{z_{\mathrm{eq}}}-1\right)~,
\end{align}
where $z_{eq}$ is the redshift at matter-radiation equality, $z_f$ is the redshift of PBH formation, $H_0$ is the Hubble constant, and $\Omega_r$ is today's radiation density. The primordial black holes form when their physical mass is related to the Hubble mass by $m = \gamma m_H$, where $\gamma$ is a numerical factor somewhat smaller than $1$ which accounts for the details of the black hole collapse. The favoured value of $\gamma$ seems to be around $0.2$ but could possibly be as low as $10^{-4}$~\cite{Carr:2009jm}. Then the redshift at formation is given by:

\begin{align}
    z_\mathrm{f}(m) = \left(\frac{2GmH_0\sqrt{\Omega_r}}{\gamma}\right)^{-1/2}~.
\end{align}
We can safely assume that the epoch of formation is much earlier than matter-radiation equality, which allows us to solve for the critical mass,

 \begin{align}
     m_* \sim 9.6 \times 10^{-13}~ \msun~ \bigg(\frac{\gamma}{0.2}\bigg)^{\frac17}\bigg(\frac{h}{0.67}\bigg)^{-\frac37}\bigg(\frac{\Omega_{\rm{r}}}{5.4\times10^{-5}}\bigg)^{-\frac{3}{14}}~,
 \end{align}
where we have used the \textit{Planck} 2015 values~\cite{Ade:2015xua} and the standard $H_0 = h\times100$ km/s.Mpc for the cosmological parameters.

\subsection{Constraints on Thakurta PBH dark matter}
As discussed earlier, this critical mass represents the upper bound of a stability constraint on PBH dark matter candidates of mass smaller than $m_*$. A black hole which formed at a mass very slightly higher than this value would still leave behind remnants which could today comprise some of the dark matter. However, such a theory would depend extremely sensitively on the formation mass; a very tiny deviation can change the mass of the remnants today by orders of magnitude. This sensitivity is smaller than the precision with which we can conduct these calculations, so we can treat it as a `stability constraint' on the theory. 

This is the same kind of constraint as the standard Schwarzschild constraint on PBHs of mass smaller than $10^{-19}~\msun$. However, in the usual case, Hawking radiation today from the remnants would still be observable and strong. In our case, the remnants left behind could be in the asteroid-mass range, where there would not be strong Hawking radiation signatures today, so we must carefully describe this stability constraint in more depth than it is often discussed.

There is also the fact that the mass spectrum of PBHs will never be truly monochromatic~\cite{Carr:2017jsz,Gow:2020cou,Niemeyer:1997mt,Yokoyama:1998xd,Kuhnel:2015vtw}. Even an idealized physically realistic scenario will include some spread around the peak mass (and the mass spectrum may be more complicated than that anyway). This variance in the PBH initial masses would make it extremely difficult to populate a sizeable fraction of the dark matter today with a specific mass of PBH in the asteroid-mass range. To demonstrate this qualitatively, we've included our new constraint in Fig.~\ref{fig:pbhconstraints} with a gradient to indicate a higher level of uncertainty at small fractions of the dark matter population. Quantitatively including the effects of extended mass distributions to the conclusions here goes beyond the scope of this work, but would be an interesting avenue for future exploration.

\begin{figure}[t]
    \centering
    \includegraphics[width=0.8\textwidth]{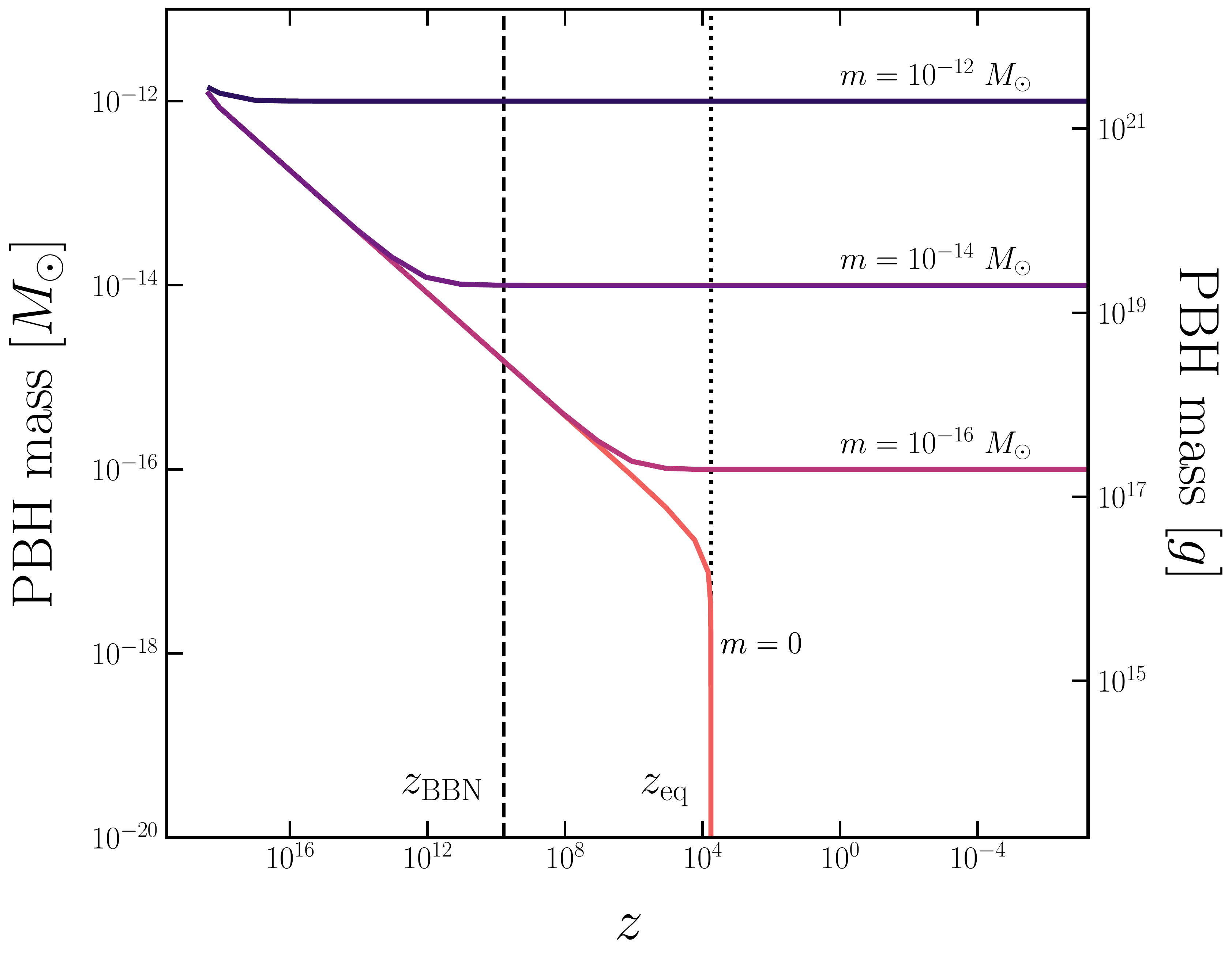}
    \caption{\label{fig:pbh_hawking} PBH mass evolutions, numerically evaluated from Eq.~(\ref{dmdt}), starting from extremely similar initial conditions. The topmost line barely evaporates, while the bottom orange line has fully evaporated by matter-radiation equality. One can see that the difference in initial mass is miniscule between these two scenarios; the bottom three curves start with initial masses which are the same to nine decimal places, while the topmost curve is about $1.15$ times more massive than the others. Here we assume that the black holes are described by the Thakurta solution until $z_{\rm eq}$, which is an extremely conservative decoupling time. After $z_{\rm eq}$ they then evaporate as they would in the Schwarzschild (or Einstein-Strauss) solution. We can see from the sensitivity to initial conditions here that populating all of the dark matter today with black holes below the critical mass $m_*$ is not a stable solution. Additionally, we can see that during the epoch of BBN, there will be significant energy ejection for PBHs which end up in much of the asteroid-mass range today; this would pose an additional strong constraint on them as dark matter candidates.}
\end{figure}

During the efficient evaporation of Thakurta black holes in the early universe, there would be a significant amount of energy and particle injection into the cosmological bath. Big Bang Nucleosynthesis (BBN) calculations place tight constraints on early universe processes at a redshift of $z_{\mathrm{BBN}} \sim 6\times10^9$, comprising some of the tightest bounds on evaporating black holes of very small initial masses~\cite{Miyama:1978mp,vainer1978SvA....22..138V,Khlopov:1980mg,Kohri:1999ex,Kawasaki:2004qu,Carr:2009jm}. In the Thakurta case, it is not absurd to imagine that the particle injection from much larger PBHs might be similar in magnitude and substance to the particle injection from these very small Schwarzschild PBHs, and so would place similarly tight constraints on the fraction of the universe's energy density in black holes. Such a calculation again goes beyond the scope of this work, but is worthy of future investigation.

\section{Conclusion}\label{sec:conc}

The constraints on PBH dark matter depend heavily on the modelling of these black holes as cosmologically embedded objects in the early universe. We explored one such cosmological PBH solution-- the Thakurta solution-- which stands out by its generality, simplicity, lack of pathologies, and applicability in the radiation-dominated early universe. The Thakurta black hole possesses a time-varying apparent horizon which grows proportionally to the scale factor. Because of this, it produces significantly more Hawking radiation and so possesses a much shorter lifetime than that of a Schwarzschild black hole of the same mass. As a result, we are able to rule out almost the entirety of the asteroid-mass range of PBHs as viable dark matter candidates. Even by quite conservative assumptions about the decoupling time of Thakurta black holes, PBHs of masses smaller than the critical mass $m_*\sim10^{-12}~\msun$, or around $m_*\sim 10^{21}~g$, evaporate with a lifetime shorter than the age of the universe. This constitutes a bound on PBHs forming dark matter \textit{today} which are lighter than this critical mass, since a theory which would leave such remnants today is extremely sensitive to the PBH formation mass and so should be considered unstable.

Our conclusion here would constitute a relatively dramatic change in the landscape of PBH dark matter bounds\footnote{It was previously estimated that the odds of navigating the asteroid field was $3,720$~to~$1$~\cite{star}.}. However, even if the Thakurta solution is not truly the most adequate description of early universe PBHs, there are important and generic consequences of our work here. Just the choice of black hole metric alone leads to significant and wide-ranging phenomenological impacts on the viability of PBH dark matter theories. Such a choice is far from trivial and should be taken seriously, since there are unambiguous problems with the previous Schwarzschild description.

The consequences of our arguments here may extend beyond the bounds of this work, which constitutes the preliminary calculations of PBH lifetimes in the Thakurta metric. Other early-universe constraints such as those from BBN, or even from impacts on the cosmic microwave background~\cite{zeldovich1977SvAL....3..110Z,sunyaev2009AN....330..657S,huPhysRevLett.70.2661,Tashiro:2008sf} or cosmic ray fluxes~\cite{page1976ApJ...206....1P,macgibbon1991ApJ...371..447M,CARR1998141}, would be worthy of investigating with different cosmological PBH metrics. In our previous work~\cite{boehm2020eliminating}, we tantalizingly re-opened the possibility that the dark matter could entirely be comprised of LIGO-mass PBHs. It is clear from our work so far on Thakurta black holes that the entire landscape of PBH dark matter sensitively depends on the choice of PBH metric and so many of the calculations and results that formerly seemed `set in stone' now deserve new and more detailed investigations.

\section*{Acknowledgements}
I would like to thank Archil Kobakhidze, C\'eline B\oe hm, Ciaran O'Hare, Markus Mosbech, and Joseph Allingham for many useful conversations, insight, and significant support throughout the research and writing of this paper.

\bibliography{bib.bib}

\providecommand{\href}[2]{#2}\begingroup\raggedright\begin{thebibliography}{10}

\bibitem{pbh}
Y.~B. Zel'dovich and I.~D. Novikov, \emph{{The hypothesis of cores retarded
  during expansion and the hot cosmological model}}, {\emph{Sov. Astron.}
  {\bfseries 10} (1966) 602}.

\bibitem{Hawking:1971ei}
S.~Hawking, \emph{{Gravitationally collapsed objects of very low mass}},
  {\emph{Mon. Not. Roy. Astron. Soc.} {\bfseries 152} (1971) 75}.

\bibitem{Carr:1974nx}
B.~J. Carr and S.~Hawking, \emph{{Black holes in the early Universe}},
  {\emph{Mon. Not. Roy. Astron. Soc.} {\bfseries 168} (1974) 399}.

\bibitem{Chapline:1975ojl}
G.~F. Chapline, \emph{{Cosmological effects of primordial black holes}},
  \href{https://doi.org/10.1038/253251a0}{\emph{Nature} {\bfseries 253} (1975)
  251}.

\bibitem{Hawking:1974sw}
S.~W. Hawking, \emph{{Particle Creation by Black Holes}},
  \href{https://doi.org/10.1007/BF02345020}{\emph{Commun. Math. Phys.}
  {\bfseries 43} (1975) 199}. [Erratum: Commun.Math.Phys. 46, 206 (1976)].

\bibitem{Carr:2009jm}
B.~J. Carr, K.~Kohri, Y.~Sendouda and J.~Yokoyama, \emph{{New cosmological
  constraints on primordial black holes}},
  \href{https://doi.org/10.1103/PhysRevD.81.104019}{\emph{Phys. Rev. D}
  {\bfseries 81} (2010) 104019}
  [\href{https://arxiv.org/abs/0912.5297}{{\ttfamily 0912.5297}}].

\bibitem{Niikura:2017zjd}
H.~Niikura et~al., \emph{{Microlensing constraints on primordial black holes
  with Subaru/HSC Andromeda observations}},
  \href{https://doi.org/10.1038/s41550-019-0723-1}{\emph{Nature Astron.}
  {\bfseries 3} (2019) 524} [\href{https://arxiv.org/abs/1701.02151}{{\ttfamily
  1701.02151}}].

\bibitem{Montero_Camacho_2019}
P.~Montero-Camacho, X.~Fang, G.~Vasquez, M.~Silva and C.~M. Hirata,
  \emph{Revisiting constraints on asteroid-mass primordial black holes as dark
  matter candidates},
  \href{https://doi.org/10.1088/1475-7516/2019/08/031}{\emph{Journal of
  Cosmology and Astroparticle Physics} {\bfseries 2019} (2019) 031}.

\bibitem{Green:2020jor}
A.~M. Green and B.~J. Kavanagh, \emph{{Primordial Black Holes as a dark matter
  candidate}},  \href{https://arxiv.org/abs/2007.10722}{{\ttfamily
  2007.10722}}.

\bibitem{Carr:2020gox}
B.~Carr, K.~Kohri, Y.~Sendouda and J.~Yokoyama, \emph{{Constraints on
  Primordial Black Holes}},  \href{https://arxiv.org/abs/2002.12778}{{\ttfamily
  2002.12778}}.

\bibitem{McVittie:1933zz}
G.~McVittie, \emph{{The mass-particle in an expanding universe}},
  \href{https://doi.org/10.1093/mnras/93.5.325}{\emph{Mon. Not. Roy. Astron.
  Soc.} {\bfseries 93} (1933) 325}.

\bibitem{Einstein:1945id}
A.~Einstein and E.~G. Straus, \emph{{The influence of the expansion of space on
  the gravitation fields surrounding the individual stars}},
  \href{https://doi.org/10.1103/RevModPhys.17.120}{\emph{Rev. Mod. Phys.}
  {\bfseries 17} (1945) 120}.

\bibitem{Faraoni:2018xwo}
V.~Faraoni, \emph{{Embedding black holes and other inhomogeneities in the
  universe in various theories of gravity: a short review}},
  \href{https://doi.org/10.3390/universe4100109}{\emph{Universe} {\bfseries 4}
  (2018) 109} [\href{https://arxiv.org/abs/1810.04667}{{\ttfamily
  1810.04667}}].

\bibitem{Thakurta}
G.~S.~N. Thakurta, \emph{{Kerr metric in an expanding universe}}, {\emph{Indian
  J. Phys.} {\bfseries 55B} (1981) 304}.

\bibitem{boehm2020eliminating}
C.~Boehm, A.~Kobakhidze, C.~A.~J. O'Hare, Z.~S.~C. Picker and M.~Sakellariadou,
  \emph{{Eliminating the LIGO bounds on primordial black hole dark matter}},
  \href{https://arxiv.org/abs/2008.10743}{{\ttfamily 2008.10743}}.

\bibitem{Abbott:2016blz}
{\scshape LIGO Scientific, Virgo} Collaboration, B.~Abbott et~al.,
  \emph{{Observation of Gravitational Waves from a Binary Black Hole Merger}},
  \href{https://doi.org/10.1103/PhysRevLett.116.061102}{\emph{Phys. Rev. Lett.}
  {\bfseries 116} (2016) 061102}
  [\href{https://arxiv.org/abs/1602.03837}{{\ttfamily 1602.03837}}].

\bibitem{Nakamura:1997sm}
T.~Nakamura, M.~Sasaki, T.~Tanaka and K.~S. Thorne, \emph{{Gravitational waves
  from coalescing black hole MACHO binaries}},
  \href{https://doi.org/10.1086/310886}{\emph{Astrophys. J. Lett.} {\bfseries
  487} (1997) L139} [\href{https://arxiv.org/abs/astro-ph/9708060}{{\ttfamily
  astro-ph/9708060}}].

\bibitem{Ali-Haimoud:2017rtz}
Y.~Ali-Haïmoud, E.~D. Kovetz and M.~Kamionkowski, \emph{{Merger rate of
  primordial black-hole binaries}},
  \href{https://doi.org/10.1103/PhysRevD.96.123523}{\emph{Phys. Rev. D}
  {\bfseries 96} (2017) 123523}
  [\href{https://arxiv.org/abs/1709.06576}{{\ttfamily 1709.06576}}].

\bibitem{bradley_j_kavanagh_2019_3538999}
B.~J. Kavanagh, \emph{bradkav/pbhbounds: Release version},  Nov., 2019.
\newblock 10.5281/zenodo.3538999.

\bibitem{Kodama:1979vn}
H.~Kodama, \emph{{Conserved Energy Flux for the Spherically Symmetric System
  and the Back Reaction Problem in the Black Hole Evaporation}},
  \href{https://doi.org/10.1143/PTP.63.1217}{\emph{Prog. Theor. Phys.}
  {\bfseries 63} (1980) 1217}.

\bibitem{Abreu:2010ru}
G.~Abreu and M.~Visser, \emph{{Kodama time: Geometrically preferred foliations
  of spherically symmetric spacetimes}},
  \href{https://doi.org/10.1103/PhysRevD.82.044027}{\emph{Phys. Rev. D}
  {\bfseries 82} (2010) 044027}
  [\href{https://arxiv.org/abs/1004.1456}{{\ttfamily 1004.1456}}].

\bibitem{carrera}
M.~Carrera and D.~Giulini, \emph{Influence of global cosmological expansion on
  local dynamics and kinematics},
  \href{https://doi.org/10.1103/RevModPhys.82.169}{\emph{Rev. Mod. Phys.}
  {\bfseries 82} (2010) 169}.

\bibitem{Tolman:1934za}
R.~C. Tolman, \emph{{Effect of imhomogeneity on cosmological models}},
  \href{https://doi.org/10.1073/pnas.20.3.169}{\emph{Proc. Nat. Acad. Sci.}
  {\bfseries 20} (1934) 169}.

\bibitem{Bondi:1947fta}
H.~Bondi, \emph{{Spherically symmetrical models in general relativity}},
  \href{https://doi.org/10.1093/mnras/107.5-6.410}{\emph{Mon. Not. Roy. Astron.
  Soc.} {\bfseries 107} (1947) 410}.

\bibitem{Joshi:2014gea}
P.~S. Joshi and D.~Malafarina, \emph{{All black holes in
  Lema\^\i{}tre\textendash{}Tolman\textendash{}Bondi inhomogeneous dust
  collapse}},
  \href{https://doi.org/10.1088/0264-9381/32/14/145004}{\emph{Class. Quant.
  Grav.} {\bfseries 32} (2015) 145004}
  [\href{https://arxiv.org/abs/1405.1146}{{\ttfamily 1405.1146}}].

\bibitem{Sultana:2005tp}
J.~Sultana and C.~C. Dyer, \emph{{Cosmological black holes: A black hole in the
  Einstein-de Sitter universe}},
  \href{https://doi.org/10.1007/s10714-005-0119-7}{\emph{Gen. Rel. Grav.}
  {\bfseries 37} (2005) 1347}.

\bibitem{Faraoni:2007es}
V.~Faraoni and A.~Jacques, \emph{{Cosmological expansion and local physics}},
  \href{https://doi.org/10.1103/PhysRevD.76.063510}{\emph{Phys. Rev. D}
  {\bfseries 76} (2007) 063510}
  [\href{https://arxiv.org/abs/0707.1350}{{\ttfamily 0707.1350}}].

\bibitem{clifton10.1111/j.1365-2966.2005.08831.x}
T.~Clifton, D.~F. Mota and J.~D. Barrow, \emph{{Inhomogeneous gravity}},
  \href{https://doi.org/10.1111/j.1365-2966.2005.08831.x}{\emph{Monthly Notices
  of the Royal Astronomical Society} {\bfseries 358} (2005) 601}.

\bibitem{Arnowitt:1959ah}
R.~L. Arnowitt, S.~Deser and C.~W. Misner, \emph{{Dynamical Structure and
  Definition of Energy in General Relativity}},
  \href{https://doi.org/10.1103/PhysRev.116.1322}{\emph{Phys. Rev.} {\bfseries
  116} (1959) 1322}.

\bibitem{Misner:1964je}
C.~W. Misner and D.~H. Sharp, \emph{{Relativistic equations for adiabatic,
  spherically symmetric gravitational collapse}},
  \href{https://doi.org/10.1103/PhysRev.136.B571}{\emph{Phys. Rev.} {\bfseries
  136} (1964) B571}.

\bibitem{boothdoi:10.1139/p05-063}
I.~Booth, \emph{Black-hole boundaries},
  \href{https://doi.org/10.1139/p05-063}{\emph{Canadian Journal of Physics}
  {\bfseries 83} (2005) 1073}.

\bibitem{Ade:2015xua}
{\scshape Planck} Collaboration, P.~Ade et~al., \emph{{Planck 2015 results.
  XIII. Cosmological parameters}},
  \href{https://doi.org/10.1051/0004-6361/201525830}{\emph{Astron. Astrophys.}
  {\bfseries 594} (2016) A13}
  [\href{https://arxiv.org/abs/1502.01589}{{\ttfamily 1502.01589}}].

\bibitem{Carr:2017jsz}
B.~Carr, M.~Raidal, T.~Tenkanen, V.~Vaskonen and H.~Veerm\"ae,
  \emph{{Primordial black hole constraints for extended mass functions}},
  \href{https://doi.org/10.1103/PhysRevD.96.023514}{\emph{Phys. Rev. D}
  {\bfseries 96} (2017) 023514}
  [\href{https://arxiv.org/abs/1705.05567}{{\ttfamily 1705.05567}}].

\bibitem{Gow:2020cou}
A.~D. Gow, C.~T. Byrnes and A.~Hall, \emph{{Primordial black holes from narrow
  peaks and the skew-lognormal distribution}},
  \href{https://arxiv.org/abs/2009.03204}{{\ttfamily 2009.03204}}.

\bibitem{Niemeyer:1997mt}
J.~C. Niemeyer and K.~Jedamzik, \emph{{Near-critical gravitational collapse and
  the initial mass function of primordial black holes}},
  \href{https://doi.org/10.1103/PhysRevLett.80.5481}{\emph{Phys. Rev. Lett.}
  {\bfseries 80} (1998) 5481}
  [\href{https://arxiv.org/abs/astro-ph/9709072}{{\ttfamily
  astro-ph/9709072}}].

\bibitem{Yokoyama:1998xd}
J.~Yokoyama, \emph{{Cosmological constraints on primordial black holes produced
  in the near critical gravitational collapse}},
  \href{https://doi.org/10.1103/PhysRevD.58.107502}{\emph{Phys. Rev. D}
  {\bfseries 58} (1998) 107502}
  [\href{https://arxiv.org/abs/gr-qc/9804041}{{\ttfamily gr-qc/9804041}}].

\bibitem{Kuhnel:2015vtw}
F.~K\"uhnel, C.~Rampf and M.~Sandstad, \emph{{Effects of Critical Collapse on
  Primordial Black-Hole Mass Spectra}},
  \href{https://doi.org/10.1140/epjc/s10052-016-3945-8}{\emph{Eur. Phys. J. C}
  {\bfseries 76} (2016) 93} [\href{https://arxiv.org/abs/1512.00488}{{\ttfamily
  1512.00488}}].

\bibitem{Miyama:1978mp}
S.~Miyama and K.~Sato, \emph{{The Upper Bound of the Number Density of
  Primordial Black Holes From the Big Bang Nucleosynthesis}},
  \href{https://doi.org/10.1143/PTP.59.1012}{\emph{Prog. Theor. Phys.}
  {\bfseries 59} (1978) 1012}.

\bibitem{vainer1978SvA....22..138V}
B.~V. {Vainer} and P.~D. {Naselskii}, \emph{{Cosmological implications of the
  process of primordial black hole evaporation}}, {\emph{sovast} {\bfseries 22}
  (1978) 138}.

\bibitem{Khlopov:1980mg}
M.~Y. Khlopov and A.~G. Polnarev, \emph{{PRIMORDIAL BLACK HOLES AS A
  COSMOLOGICAL TEST OF GRAND UNIFICATION}},
  \href{https://doi.org/10.1016/0370-2693(80)90624-3}{\emph{Phys. Lett. B}
  {\bfseries 97} (1980) 383}.

\bibitem{Kohri:1999ex}
K.~Kohri and J.~Yokoyama, \emph{{Primordial black holes and primordial
  nucleosynthesis. 1. Effects of hadron injection from low mass holes}},
  \href{https://doi.org/10.1103/PhysRevD.61.023501}{\emph{Phys. Rev. D}
  {\bfseries 61} (2000) 023501}
  [\href{https://arxiv.org/abs/astro-ph/9908160}{{\ttfamily
  astro-ph/9908160}}].

\bibitem{Kawasaki:2004qu}
M.~Kawasaki, K.~Kohri and T.~Moroi, \emph{{Big-Bang nucleosynthesis and
  hadronic decay of long-lived massive particles}},
  \href{https://doi.org/10.1103/PhysRevD.71.083502}{\emph{Phys. Rev. D}
  {\bfseries 71} (2005) 083502}
  [\href{https://arxiv.org/abs/astro-ph/0408426}{{\ttfamily
  astro-ph/0408426}}].

\bibitem{star}
G.~Lucas, \emph{Star {W}ars: Episode {V} -- {T}he {E}mpire {S}trikes {B}ack},
  20th Century Fox, May, 1980.

\bibitem{zeldovich1977SvAL....3..110Z}
I.~B. {Zeldovich}, A.~A. {Starobinskii}, M.~I. {Khlopov} and V.~M.
  {Chechetkin}, \emph{{Primordial black holes and the deuterium problem}},
  {\emph{Soviet Astronomy Letters} {\bfseries 3} (1977) 110}.

\bibitem{sunyaev2009AN....330..657S}
R.~A. {Sunyaev} and J.~{Chluba}, \emph{{Signals from the epoch of cosmological
  recombination (Karl Schwarzschild Award Lecture 2008)}},
  \href{https://doi.org/10.1002/asna.200911237}{\emph{Astronomische
  Nachrichten} {\bfseries 330} (2009) 657}
  [\href{https://arxiv.org/abs/0908.0435}{{\ttfamily 0908.0435}}].

\bibitem{huPhysRevLett.70.2661}
W.~Hu and J.~Silk, \emph{Thermalization constraints and spectral distortions
  for massive unstable relic particles},
  \href{https://doi.org/10.1103/PhysRevLett.70.2661}{\emph{Phys. Rev. Lett.}
  {\bfseries 70} (1993) 2661}.

\bibitem{Tashiro:2008sf}
H.~Tashiro and N.~Sugiyama, \emph{{Constraints on Primordial Black Holes by
  Distortions of Cosmic Microwave Background}},
  \href{https://doi.org/10.1103/PhysRevD.78.023004}{\emph{Phys. Rev. D}
  {\bfseries 78} (2008) 023004}
  [\href{https://arxiv.org/abs/0801.3172}{{\ttfamily 0801.3172}}].

\bibitem{page1976ApJ...206....1P}
D.~N. {Page} and S.~W. {Hawking}, \emph{{Gamma rays from primordial black
  holes.}}, \href{https://doi.org/10.1086/154350}{\emph{apj} {\bfseries 206}
  (1976) 1}.

\bibitem{macgibbon1991ApJ...371..447M}
J.~H. {MacGibbon} and B.~J. {Carr}, \emph{{Cosmic Rays from Primordial Black
  Holes}}, \href{https://doi.org/10.1086/169909}{\emph{apj} {\bfseries 371}
  (1991) 447}.

\bibitem{CARR1998141}
B.~Carr and J.~MacGibbon, \emph{Cosmic rays from primordial black holes and
  constraints on the early universe},
  \href{https://doi.org/https://doi.org/10.1016/S0370-1573(98)00039-8}{\emph{Physics
  Reports} {\bfseries 307} (1998) 141}.

\end{thebibliography}\endgroup
\bibliographystyle{bibi}

\end{document}